\documentclass[12pt,centerh1]{article}

\usepackage{graphicx}

\usepackage{natbib,color}
\usepackage{epsfig,soul}
\usepackage{amsfonts} 
\usepackage{amsmath}
\usepackage{amssymb,longtable,enumitem}
\usepackage{setspace}

\newcommand{\vecx}{\mathbf{x}}
\newcommand{\vecy}{\mathbf{y}}

\newcommand{\varthet}{\mbox{$\boldsymbol{\vartheta}$}}

\newcommand{\matd}{\mathbf{D}}
\newcommand{\mata}{\mathbf{A}}
\newcommand{\ident}{\mathbf{I}}
\newcommand{\tr}{\text{tr}}

\newcommand{\del}{\mathbf{A}}
\newcommand{\gam}{\mathbf{D}}
\newcommand{\Mu}{\boldsymbol{\mu}}
\newcommand{\vecmu}{\mbox{\boldmath$\mu$}}

\newcommand{\vectheta}{\mbox{$\boldsymbol{\theta}$}}

\newcommand{\matsig}{\mathbf{\Sigma}}

\newcommand{\matt}{\mathbf{T}}

\newcommand{\ex}{\mathbb{E}}

\date{}
\LTcapwidth=\textwidth
\begin{document}

\title{Parameter Estimation and Mixture Model Selection Within a Family Setting}

\author{Sanjeena Subedi\footnote{Department of Mathematical Sciences, Binghamton University, State University of New York, 4400 Vestal Parkway East, Binghamton, NY, USA 13902. e: sdang@binghamton.edu} \and Paul D.\ McNicholas \footnote{Department of Mathematics \& Statistics, McMaster University, 1280 Main St.~W., Hamilton, ON, Canada L8S 4K1. e: mcnicholas@math.mcmaster.ca}}

\maketitle

\begin{abstract}
Mixture model-based clustering has become an increasingly popular data analysis technique since its introduction over fifty years ago, and is now commonly utilized within a family setting. Families of mixture models arise when the component parameters, usually the component covariance (or scale) matrices, are decomposed and a number of constraints are imposed. Within the family setting, model selection involves choosing the member of the family, i.e., the appropriate covariance structure, in addition to the number of mixture components. To date, the Bayesian information criterion (BIC) has proved most effective for model selection, and the expectation-maximization (EM) algorithm is usually used for parameter estimation. In fact, this EM-BIC rubric has virtually monopolized the literature on families of mixture models. Deviating from this rubric, variational Bayes approximations are developed for parameter estimation and the deviance information criterion for model selection. The variational Bayes approach provides an alternate framework for parameter estimation by constructing a tight lower bound on the complex marginal likelihood and maximizing this lower bound by minimizing the associated Kullback-Leibler divergence. This approach is taken on the most commonly used family of Gaussian mixture models, and real and simulated data are used to compare the new approach to the EM-BIC rubric.\end{abstract}


\section{Introduction}

Most early clustering algorithms were based on heuristic approaches and some such methods, including hierarchical agglomerative clustering and $k$-means clustering \citep{macqueen67,hartigan79}, are still widely used. The use of mixture models to account for population heterogeneity has been very well established for over a century \citep[e.g.,][]{pearson1894}, but it was the 1960s before mixture models were used for clustering \citep{wolfe65,hasselblad66,day69}. Because of the lack of suitable computing equipment, it was much later before the use of mixture models \citep[e.g.,][]{banfield93,celeux95} and, more generally, probability models \citep[e.g.,][]{bock96,bock98b,bock98a} for clustering became commonplace. Since the turn of the century, the use of mixture models for clustering has burgeoned into a popular subfield of cluster analysis and recent examples include: \cite{franczak14}, \cite{vrbik14}, \cite{murray14b,murray14a}, \cite{lee14}, \cite{lin14}, \cite{subedi2015}, \cite{morris15}, \cite{ohagan16}, \cite{dang2015}, \cite{lin16}, \cite{lee2016}, \cite{dang2017}, \cite{cheam2017}, \cite{melnykov18}, \cite{zhu18}, \cite{gallaugher19}, \cite{tortora19}, \cite{biernacki19}, \cite{murray19}, \cite{morris19}, and \cite{punzo20}. The reader may consult \cite{bouveyron14} and \cite{mcnicholas16b} for relatively recent reviews of model-based clustering work.

 A $d$-dimensional random vector $\mathbf{Y}$ is said to arise from a parametric finite mixture distribution if, for all $\vecy \subset \mathbf{Y}$, we can write its density as $$f(\vecy\mid\varthet)= \sum_{g=1}^G \rho_g p_g(\vecy\mid\vectheta_g),$$
where $\rho_g>0$ such that $\sum_{i=1}^G \rho_{g} =1$ are the mixing proportions, $p_g(\vecy\mid\vectheta_g)$ are component densities, and $\varthet=(\rho_1,\ldots,\rho_G,\vectheta_1,\ldots,\vectheta_G)$ is the vector of parameters. 
When the component parameters $\vectheta_1,\ldots,\vectheta_G$ are decomposed and constraints are imposed on the resulting decompositions, the result is a family of mixture models. Typically, each component probability density is of the same type and, when they are Gaussian, the density function is $$f(\vecy\mid\varthet)= \sum_{g=1}^G \rho_g\phi_d(\vecy\mid\vecmu_g,\matsig_g),$$ where $\phi_d(\vecy\mid\vecmu_g,\matsig_g)$ is the $d$-dimensional Gaussian density with mean $\vecmu_g$ and covariance $\matsig_g$, and  the likelihood is
$$\mathcal{L}(\boldsymbol{\vartheta} \mid \vecy_1, \ldots, \vecy_n) = \prod_{i=1}^{n} \sum_{g=1}^G\rho_g\phi_d(\vecy_i\mid \boldsymbol{\mu}_g,\matsig_g),$$
where $\boldsymbol{\vartheta}$ denotes the model parameters. In Gaussian families, it is usually the component covariance matrices $\matsig_1,\ldots,\matsig_G$ that are decomposed (see Section~\ref{sec:method}).

The expectation-maximization (EM) algorithm \citep{dempster77} is often used for mixture model parameter estimation but its efficacy is questionable. As discussed by \cite{titterington85} and others, the nature of the mixture likelihood surface leaves the EM algorithm open to failure. Although this weakness can be mitigated by using multiple re-starts, there is no way to completely overcome it. Besides its heavy reliance on starting values, convergence of the EM algorithm can be very slow. When families of mixture models are used, the EM algorithm approach must be employed in conjunction with a model selection criterion to select the member of the family and, in many cases, the number of components. There are many model selection criteria to choose from, such as the Bayesian information criterion \citep[BIC;][]{schwartz78}, the integrated completed likelihood \citep[ICL;][]{Biernacki00}, and the Akaike information criterion \citep[AIC;][]{akaike74}. All of these model selection criteria have some merit and various shortcomings, but the BIC remains by far the most popular \citep[][Chp.~2]{mcnicholas16a}. 
There has been interest in the use of Bayesian approaches to mixture model parameter estimation, via Markov chain Monte Carlo (MCMC) methods \citep[e.g.,][]{diebolt94,richardson97,bensmail95,stephens97,stephens00,casella02}; however, difficulties have been encountered with, \textit{inter alia}, computational overhead and convergence \citep[see][]{celeux00,jasra05}. Variational Bayes approximations present an alternative to MCMC algorithms for mixture model parameter estimation and are gaining popularity due to their fast and deterministic nature \citep[see][]{jordan99,corduneanu01,ueda02,mcgrory07,mcgrory09a,mcgrory09b,subedi2014}. 

With the use of a computationally convenient approximating density in place of a more complex `true' posterior density, the variational algorithm overcomes the hurdles of MCMC sampling. 
For observed data $\vecy$, the joint conditional distribution of parameters $\boldsymbol{\theta}$ and missing data $\mathbf{z}$ are approximated by using another computationally convenient distribution $q(\mathbf{\boldsymbol{\theta},z})$. This distribution $q(\mathbf{\boldsymbol{\theta},z})$ is obtained by minimizing the Kullback-Leibler (KL) divergence between the true and the approximating densities, where
\begin{equation*}
\text{KL} (q(\mathbf{\boldsymbol{\theta},z})\mid p(\mathbf{\boldsymbol{\theta},z\mid \mathbf{y}})) = \int_{\boldsymbol{\Theta}} \sum_{z} q(\mathbf{\boldsymbol{\theta},z}) \log\left\{\frac{q(\boldsymbol{\theta},\mathbf{z})}{p(\mathbf{\boldsymbol{\theta},z\mid y})}\right\}  d\boldsymbol{\theta}.
\end{equation*}
The approximating density is restricted to have a factorized form for computational convenience, so that $q(\boldsymbol{\theta},\mathbf{z})=q_{\theta}(\boldsymbol{\theta})q_{z}(\mathbf{z})$. Upon choosing a conjugate prior, the appropriate hyper-parameters of the approximating density $q_{\theta}(\boldsymbol{\theta})$ can be obtained by solving a set of coupled non-linear equations.

The variational Bayes algorithm is initialized with more components than expected. As the algorithm iterates, if two components have  similar parameters then one component dominates the other causing the dominated component's weighting to be zero. If a component's weight becomes sufficiently small, less than or equal to two observations in our analyses, the component is removed from consideration. Therefore, the variational Bayes approach allows for simultaneous parameter estimation and selection of the number of components.

\section{Methodology}\label{sec:method}
\subsection{Introducing Parsimony}
If $d$-dimensional data $\mathbf{y}_1,\ldots,\mathbf{y}_n$ arise from a finite mixture of Gaussian distributions, then the log-likelihood is
\begin{equation*}
\log p(\mathbf{y}_1,\ldots,\mathbf{y}_n \mid \vectheta) = \ \sum_{i=1}^n \log\Bigg[ \sum_{g=1}^G \rho_g\frac{|\matsig^{-1}_g|}{2\pi^\frac{d}{2}}\exp\left\{\frac{1}{2}(\vecy_i-\Mu_g)'\matsig^{-1}_g(\vecy_i-\Mu_g)\right\}\Bigg].
\end{equation*}
The number of parameters in the component covariance matrices of is ${Gd(d+1)}/{2}$, which is quadratic in $d$. When dealing with real data, the number of free parameters to be estimated can very easily exceed the sample size $n$ by an order of magnitude. Hence, the introduction of parsimony through the imposition of additional structure on the covariance matrices is desirable.
\cite{banfield93} exploited geometrical constraints on the covariance matrices of Gaussian distribution using the eigen-decomposition of the covariance matrices, such that $\matsig_g = \lambda_g \matd_g\mata_g\matd_g'$, where $\matd_g$ is the orthogonal matrix of eigenvectors and $\mata_g$ is a diagonal matrix proportional to the eigenvalues of $\matsig_g$, such that $|\mata_g|=1$, and $\lambda_g$ is the associated constant of proportionality. This decomposition has an advantage in terms of its interpretation, i.e., the parameter $\lambda_g$ controls the cluster volume, $\mata_g$ controls the cluster shape, and $\matd_g$ controls the cluster orientation. This allows for imposition of several constraints on the covariance matrix that have geometrical interpretation giving rise to a family of 14 models  known as Gaussian Parsimonious clustering models \citep[GPCM;][]{celeux95}; see Table~\ref{tab:GPCM}.
\begin{table}[!ht]
\centering
\caption{Nomenclature, interpretation, and covariance structure for each member of the GPCM family.}\label{tab:GPCM}
\begin{tabular*}{0.9\textwidth}{@{\extracolsep{\fill}}llllr}
\hline
 {Model} & {Volume} & {Shape} & {Orientation} & $\matsig_g$\\
\hline
EII & Equal    & Spherical &             & $\lambda \boldsymbol{I}$\\
VII & Variable & Spherical &             & $\lambda_g \boldsymbol{I}$\\
EEI & Equal    & Equal     & Axis-Aligned & $\lambda \del$\\
VEI & Variable & Equal     & Axis-Aligned & $\lambda_g \del$\\
EVI & Equal    & Variable  & Axis-Aligned & $\lambda \del_g$\\
VVI & Variable & Variable  & Axis-Aligned & $\lambda_g \del_g$\\
EEE & Equal    & Equal     & Equal        & $\lambda\gam\del\gam'$\\
VEE & Variable & Equal     & Equal        & $\lambda_g\gam\del\gam'$\\
EVE & Equal    & Variable  & Equal        & $\lambda\gam\del_g\gam'$\\
EEV & Equal    & Equal     & Variable     & $\lambda\gam_g\del\gam_g'$\\
VVE & Variable & Variable  & Equal        & $\lambda_g\gam\del_g\gam'$\\
VEV & Variable & Equal     & Variable   & $\lambda_g\gam_g\del\gam_g'$\\
EVV & Equal    & Variable  & Variable   & $\lambda\gam_g\del_g\gam_g'$\\
VVV & Variable & Variable  & Variable   & $\lambda_g\gam_g\del_g\gam_g'$\\
\hline
\end{tabular*}
\end{table}

The ${\tt mclust}$ package \citep{scrucca2016} for {\sf R} \citep{R18} implements 12 of these 14 GPCM models in an EM framework, with the MM framework of \cite{browne14a} used for the other two models (EVE and VVE). \cite{bensmail95} used Gibbs sampling to carry out Bayesian inference for eight of the GPCM models. Bayesian regularization of some of the GPCM models has been considered by \cite{fraley07}. After assigning a highly dispersed conjugate prior, they replace the maximum likelihood estimator of the group membership obtained using the EM algorithm by a maximum \textit{a~posteriori} probability (MAP) estimator. Note that $\text{MAP} (\hat{z}_{ig}) = 1$ if $g=\arg\max_h({\hat{z}_{ih}})$ and $\text{MAP}({\hat{z}_{ig}}) = 0$ otherwise, where $\hat{z}_{ig}$ denotes the \textit{a~posteriori} expected value of $Z_{ig}$ and $$z_{ig}=\begin{cases}1 & \text{if } \vecx_i \text{ belongs to component } g, \text{ and}\\
0 & \text{otherwise}.\end{cases}$$ A modified BIC using the maximum \textit{a posteriori} probability was then used for model selection. However, herein we implement 12 of those 14 GPCM models using variational Bayes approximations---conjugate priors are not available for the EVE and VVE models.
 
\subsection{Priors and Approximating Densities} 

 As suggested by \cite{mcgrory07}, the Dirichlet distribution is used as the conjugate prior for the mixing proportion, such that 
$$p(\boldsymbol{\rho})=\text{Dir}(\boldsymbol{\rho}; \alpha_1^{(0)},\ldots,\alpha_G^{(0)}),$$
where $\boldsymbol{\rho}=(\rho_1,\ldots\rho_G)$ are the mixing proportions and $\alpha_1^{(0)},\ldots,\alpha_G^{(0)}$ are the hyperparameters. Conditional on the precision matrix $\matt_g$, independent normal distributions were used as the conjugate priors for the means such that 
$$p(\Mu_1,\ldots,\Mu_G \mid\matt_1,\ldots,\matt_G)=\prod_{g=1}^{G}\phi_d(\Mu_g;\mathbf{m}^{(0)}_g,(\beta_g^{(0)}\mathbf{T}_g)^{-1}),$$
where $\{\mathbf{m}_g^{(0)}, \beta_g^{(0)}\}_{g=1}^G$ are the hyper-parameters.

\cite{fraley07} assigned priors on the parameters for the covariance matrix and its components in a Bayesian regularization application. However, we assign priors on the precision matrix with the hyperparameters given in Table~\ref{table1}. Note that it was not possible to put a suitable (i.e., determinant one) prior on the matrix $\mata_g$ for the models  EVI and VVI or on $\mata$ for models VEV and VEI; accordingly, we instead put a prior on $c_g\mata_g^{-1}$ or $c\mata^{-1}$, respectively, where $c_g$ or $c$ is a positive constant. Using the expected value of $c_g\mata_g^{-1}$ (or $c\mata^{-1}$), the expected value of $\mata_g^{-1}$ (or $\mata^{-1}$) is determined to satisfy the constraint that the determinant is 1. Because $\matd_g$ is an orthogonal matrix of eigenvectors, the Bingham matrix distribution is used as the conjugate prior for $\matd_g$. The Bingham distribution, first introduced by \cite{bingham1974}, is a probability distribution on a set of orthonormal vectors $\{\mathbf{u}: \mathbf{u}'\mathbf{u}=1\}$ and has antipodal symmetry thus making it ideal for random axes.
{\small\begin{table}[htb]
\caption{The precision parameter upon which a prior is placed, as well as the corresponding prior distribution and hyperparameters, for 12 of the 14 members of the GPCM family.}\label{table1}
\begin{tabular}{p{.07\textwidth}  p{.125\textwidth}p{.365\textwidth} p{.325\textwidth}} 
\hline
 Model & $\matsig_g$ & Precision parameter for prior & Prior and hyperparameters\\
 \hline
EII & $\lambda \ident_d$&$\lambda^{-1}$& Gamma $(a^{(0)},b^{(0)})$ \\
VII&$\lambda_g \ident_d$&$\lambda_g^{-1}$&Gamma $(a_g^{(0)},b_g^{(0)})$\\
EEI&$\lambda \mata$&$k\text{th}$ diagonal element of $(\lambda \mata)^{-1}$& Gamma $(a_k^{(0)},b_k^{(0)})$ \\
VEI &$\lambda_g \mata$&$\lambda_g^{-1}$&Gamma $(a_g^{(0)},b_g^{(0)})$\\
& &$k\text{th}$ diagonal elements of $c\mata^{-1}$& Gamma $(al_k^{(0)},be_k^{(0)})$ \\
EVI &$\lambda \mata_g$&$\lambda^{-1}$&Gamma $(a^{(0)},b^{(0)})$\\
& &$k\text{th}$ diagonal elements of $c_g\mata^{-1}_g$& Gamma $(al_{gk}^{(0)},be_{gk}^{(0)})$\\
VVI & $\lambda_g \mata_g$& $\lambda_g^{-1}$&Gamma $(a_g^{(0)},b_g^{(0)})$\\
& &$k\text{th}$ diagonal elements of $c_g\mata^{-1}_g$& Gamma $(al_{gk}^{(0)},be_{gk}^{(0)})$\\
EEE & $\lambda \matd \mata \matd'$& $\matt =(\lambda \matd \mata \matd')^{-1}$ & Wishart $(v^{(0)},\boldsymbol{\Sigma}^{(0)-1})$\\
VEE & $\lambda_g \matd \mata \matd'$ & $\lambda_g^{-1}$&Gamma $(a_g^{(0)},b_g^{(0)})$\\
& & $\matt =(\matd \mata \matd')^{-1}$ & Wishart $(v^{(0)},\boldsymbol{\Sigma}^{(0)})$\\
EEV& $\lambda \matd_g \mata \matd_g'$ &$k\text{th}$ diagonal elements of $(\lambda \mata)^{-1}$& Gamma $(a_k^{(0)},b_k^{(0)})$ \\
& & $\matd_g$ &Bingham matrix  $(\mathbf{A}_g^{(0)},\mathbf{B}_g^{(0)})$\\
VEV& $\lambda_g \matd_g\mata\matd_g'$& $\lambda_g^{-1}$& Gamma $(a_g^{(0)},b_g^{(0)})$ \\
&&$k\text{th}$ diagonal element of $c\mata^{-1}$& Gamma $(al_k^{(0)},be_k^{(0)})$ \\
& & $\matd_g$ & Bingham matrix  $(\mathbf{A}_g^{(0)},\mathbf{B}_g^{(0)})$\\
EVV& $\lambda \matd_g\mata_g\matd_g'$ &$\lambda^{-1}$& Gamma $(a^{(0)},b^{(0)})$ \\
& & $\matt_g =(\matd_g\mata_g\matd_g')^{-1}$& Wishart $(v_g^{(0)},\boldsymbol{\Sigma}_g^{(0)})$\\
VVV& $\lambda_g \matd_g\mata_g\matd_g'$& $\matt_g =(\lambda_g \matd_g\mata_g\matd_g')^{-1}$ & Wishart $(v_g^{(0)},\boldsymbol{\Sigma}_g^{(0)-1})$\\
\hline
\end{tabular}
\end{table}}

The Bingham matrix distribution \citep{gupta2000} is the matrix analogue, on the Steifel manifold, of the Bingham distribution and has been used in multivariate analysis and matrix decomposition methods \citep{hoff09}. The density of the Bingham matrix distribution, as defined by \cite{gupta2000}, is
$$p(\matd) =  b(\mathbf{A},\mathbf{B})\exp(\tr\{\mathbf{B}\matd\mathbf{A}\matd'\}) [d\matd],$$
for $\matd \in O(n,d)$, where $O(n,d)$ is the Stiefel manifold of $n\times d$ matrices, $[d\matd]$ is the unit invariant measure on $O(n,d)$, and $\mata$ and $\mathbf{B}$ are symmetric and diagonal matrices, respectively.
Samples from the Bingham matrix distribution can be obtained using the Gibbs sampling algorithm implemented in the {\sf R} package {\tt rstiefel} \citep{hoff12}.

The approximating densities that minimize the KL divergence are as follows.
For the mixing proportions,
$q_{\mathbf{\rho}}(\boldsymbol{\rho})=\text{Dir}(\boldsymbol{\rho}; \alpha_1,\ldots,\alpha_G)$,
where $\alpha_g=\alpha_g^{(0)}+\sum_{i=1}^n\hat{z}_{ig}$.
For the mean, $$q_{\Mu}(\Mu \mid\matt_1,\ldots,\matt_G)=\prod_{g=1}^{G}\phi_d(\Mu_g;\mathbf{m}_g,(\beta_g\mathbf{T}_g)^{-1}),$$
where
$\beta_g = \beta_g^{(0)}+\sum_{i=1}^n\hat{z}_{ig}$ and
$$\mathbf{m}_g =\frac{1}{\beta_g}\left({\beta_g^{(0)} \mathbf{m}_g^{(0)}+\sum_{i=1}^n\hat{z}_{ig}\mathbf{y}_i}\right).$$
The probability that the $i$th observation belongs to a group $g$ is then given by
$$\hat{z}_{ig} =\frac{\varphi_{ig}}{\sum_{j=1}^{G}\varphi_{ij}},$$
where 
\begin{equation*}\begin{split}
&\varphi_{ig}=\frac{1}{{\sum_{g=1}^{G}\varphi_{ij}}}\exp\left(\ex[\log \rho_g]+\frac{1}{2}\ex[\log|\mathbf{T}_g|]-\frac{1}{2}\mbox{tr}\left\{\ex[\mathbf{T}_g](\mathbf{y}_i-\ex[\Mu_g])(\mathbf{y}_i-\ex[\Mu_g])'+\frac{1}{\beta_g}\ident_d\right\}\right),\\
&\ex[\log(\rho_g)] =\Psi(\hat{\alpha}_g)-\Psi\left(\sum_{g=1}^G{\hat{\alpha}_g}\right),
\end{split}\end{equation*}
$\ex[\Mu_g] =\mathbf{m}_g$, and $\Psi(\cdot)$ is the digamma function. The values of $ \ex[\mathbf{T}_g]$ and $\ex[\log|\mathbf{T}_g|]$ vary depending on the model (see Table ~\ref{table3}, Appendix~\ref{app:A} for details). The posterior distribution of the parameters $\lambda_g^{-1}$ and $\mata_g$ are gamma distributions and, therefore, the expected value of $ \ex[\lambda_g^{-1}]$, $\ex[\log|\lambda_g^{-1}|]$, $ \ex[\mata_g]$, and $\ex[\log|\mata_g|]$ all have a closed form. The posterior distribution for $\matd_g\mata_g\matd_g'$ is Wishart distribution and so there is a closed form solution for $\ex[\matd_g\mata_g\matd_g']$ and $\ex[\log|\matd_g\mata_g\matd_g'|]$. The posterior distribution of the parameter $\matd_g$ is a Bingham matrix distribution (see Appendix~\ref{app:math} for details) and, hence, Monte Carlo integration was used to find the expected values of $\ex[\mathbf{T}_g]$ and $\ex[\log|\mathbf{T}_g|]$. The estimated model parameters maximize the lower bound of the marginal log-likelihood.

\subsection{Convergence}
The posterior log-likelihood of the observed data obtained using the posterior expected values of the parameters is
\begin{equation*}
\log p(\vecy_1,\ldots,\vecy_n \mid  \tilde{\vectheta}) = \sum_{i=1}^n \log \Bigg[ \sum_{g=1}^G \frac{\tilde{\rho}_g|\tilde{\matt}_g|}{2\pi^{{d}/{2}}}\exp \left\{\frac{1}{2}(\vecy_i-\tilde{\Mu}_g)'\tilde{\matt}_g(\vecy_i-\tilde{\Mu}_g)\right\}\Bigg],
\end{equation*}
where $\tilde{\vecmu}_g=\mathbf{m}_g$ and $$\tilde{\rho}_g =\frac{\alpha_g}{\sum_{j=1}^G\alpha_j}.$$ The expected precision matrix $\tilde{\matt}_g$ varies according to the model. 
Convergence of the algorithm for these models is determined using a modified Aitken acceleration criterion. 
The Aitken acceleration \citep{aitken26} is given by
\begin{equation*}
a^{(m)}=\frac{l^{(m+1)}-l^{(m)}}{l^{(m)}-l^{(m-1)}},
\end{equation*}
where $l^{(m)}$ is the value of the posterior log-likelihood at iteration $m$. Convergence can be considered to have been achieved when 
$$\big|l_\infty^{(m+1)}-l_\infty^{(m)}\big| < \epsilon,$$
where $l_\infty^{(m+1)}$ is an asymptotic estimate of the log-likelihood given by $$ l_\infty^{(m+1)} = l^{(m)}+\frac{1}{(1-a^{(m)})}(l^{(m+1)}-l^{(m-1)})$$ \citep{bohning94}. 

The VEV and EEV models utilize Gibbs sampling and Monte Carlo integration to find both the expected value of the parameter $\matt_g$ and the expectations of functions of $\matt_g$. As the Gibbs sampling chain approaches the stationary posterior distribution, the posterior log-likelihood oscillates rather than monotonically increasing at every new iteration. Hence, an alternate convergence criteria was used for these models. When the relative change in the parameter estimates from successive iterations is small, convergence is assumed. Hence, for the VEV and EEV models, the algorithm is stopped when
\begin{equation}\label{eqn:compare}
\max_i \left\{\frac{\big| \psi_i^{(m+1)}-\psi_i^{(m)}\big|}{\big| \psi_i^{(m)}\big| + \delta_1}\right\} < \delta_2,
\end{equation} 
where $\delta_1$ and $\delta_2$ are predetermined constants, $\psi_i^{(m)}$ is the estimate of the $i${th} parameter on the $m${th} iteration, and $i$ indexes over every parameter in the model---note that, for matrix- or vector-valued parameters, $\psi_i^{(m)}$ corresponds to an individual element so that $i$ indexes over all parameter elements and the comparison in \eqref{eqn:compare} is element-wise. In the analyses herein, we use $\delta_1=0.001$ and $\delta_2=0.05$ for three consecutive iterations. A detailed discussion of the convergence of Monte Carlo EM is provided in \cite{neath2013}.

 \subsection{Model Selection}
Despite the benefits of simultaneously obtaining parameter estimates along with the number of components, a model selection criterion is needed to determine the covariance structure. For the selection of the model with the best fit, the deviance information criterion \citep[DIC;][]{spiegelhalter02} is used as suggested by \cite{mcgrory07}. 
The DIC is given by 
 $$\text{DIC} = -2 \log p(\vecy_1,\ldots,\vecy_n \mid \tilde{\vectheta})+2p_D,$$ where $$2p_D \approx -2 \int q_\theta (\vectheta) \log \left\{ \frac{q_\theta (\vectheta)}{p(\vectheta)}\right\}  d\vectheta + 2 \log \left\{ \frac{q_{\theta} (\tilde{\vectheta})}{p(\tilde{\vectheta})}\right\}$$ and $\log p(\vecy_1,\ldots,\vecy_n \mid \tilde{\vectheta})$ is the posterior log-likelihood of the data.
 
Hereafter, the variational Bayes approach that uses the variational Bayes algorithms introduced herein together with the DIC to select the model (i.e., covariance structure) will be referred to as the VB-DIC approach.
  
  \subsection{Performance Assessment}
The adjusted Rand index \citep[ARI;][]{hubert85} is used to assess the performance of the clustering techniques applied in Section~\ref{sec:results}. The Rand index \citep{rand71} is based on the pairwise agreement between two partitions, e.g., predicted and true classifications. The ARI corrects the Rand index to account for agreement by chance: a value of 1 indicates perfect agreement, the expected value under random class assignment is 0, and negative values indicate a classification that is worse than would be expected by guessing.

\section{Results}\label{sec:results}

\subsection{Simulation Study 1}\label{sim1}
The VB-DIC approach is run on 50 simulated two-dimensional Gaussian data sets with three components and known mean and covariance structures $\matsig_g=\lambda_g \ident_d$ (VII, see Table~\ref{table6} for $\lambda_g$ values). For each dataset, we use five random starts for each of 12 members of the  GPCM family, and we set the maximum number of components to ten each time. For each dataset, the model with the smallest DIC is selected as the final model. A $G=3$ component model is selected on 46 out of 50 occasions with an ARI of 1 each time while a $G=4$ component model was selected, with an average ARI of 0.96 and a standard deviation of 0.044, on the other four occasions. 

A VII model is selected on 47 out of 50 occasions, with VEE and VVV models selected twice and once, respectively. When VEE and VVV are selected, the average difference between the DIC values of the model selected and VII is 1.553 with a standard deviation of 1.984 (the range of the difference was 0.470--3.049). This shows that, although the model selected was different than VII in four cases, the chosen model has similar DIC value to the VII model in each case. 
In all, there are 43 cases where a $G=3$ component VII model is selected and the true and average estimated values (with standard deviations) for $\Mu_g$ and $\lambda_g$ in these cases are given in Table~\ref{table6}---in all cases, the estimates are very close to the true values.
\begin{table}[ht]
\centering
\caption{Summary of the average and standard errors of the estimated parameters from the cases where a $G=3$ component VII model is selected in Simulation Study~1.\label{table6}}
\begin{tabular*}{0.9\textwidth}{@{\extracolsep{\fill}}llccclcc}
\hline 
&&& \multicolumn{2}{c}{$\hat{\Mu}_g$} &&  \multicolumn{2}{c}{$\hat{\lambda}_g$} \\
\cline{4-5}\cline{7-8}
$g$&$n_g$&$\Mu_g$ & Mean&St.\ Dev. & $\lambda_g$ & Mean&St.\ Dev.\\
\hline
1&250&$(-7,-7)'$& $(-6.985, -6.990)'$& $(0.081, 0.097)'$&2.2&2.199 & 0.042\\
2&100&$(-2,2)'$&$(-1.999, 1.985)'$&$(0.073, 0.066)'$&0.5&0.489 &0.101\\
3&150&$(8,0)'$&$(7.980, 0.031)'$&$(0.090, 0.090)'$&1.2&1.220 &0.130\\
\hline
\end{tabular*}
\end{table}

\begin{figure}[!ht]
\begin{center}
\includegraphics[width=\textwidth]{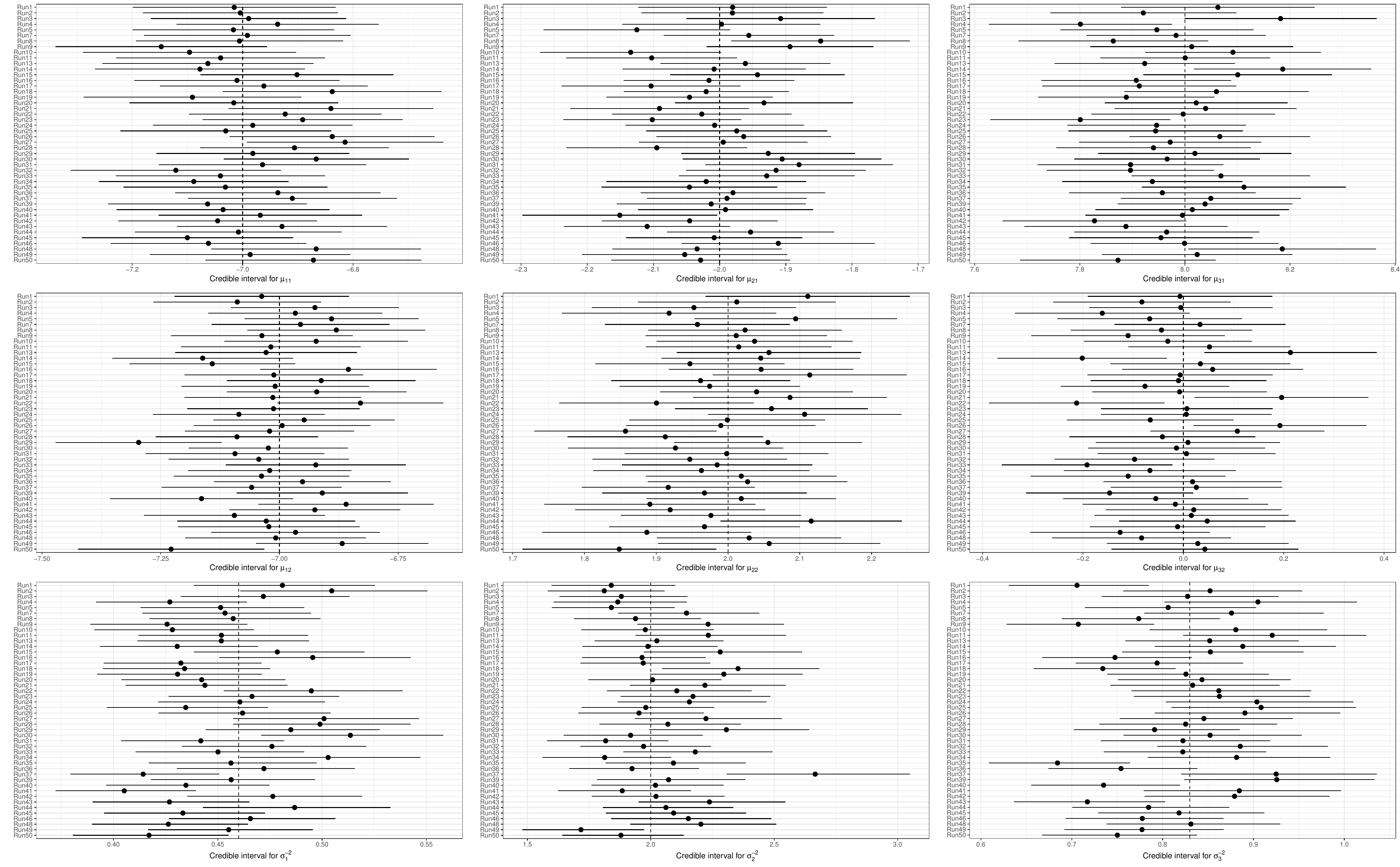} 
\caption{95\% credible intervals for the component means $\boldsymbol{\mu}_g$ (top two rows) and the component precision parameter ${1}/{\sigma^2}$ (bottom row) for the 43 runs where a $G=3$ component VII model is selected in Simulation Study~1, where vertical lines denote the values of the parameters used to generate the data.}
\label{mean_tau}
\end{center}
\vspace{-0.25in}
\end{figure}

One advantage of using a variational Bayes approach is that at every iteration, the hyperparameters of the variational posterior are updated to further minimize the Kullback-Leibler divergence between the approximate variational posterior density and the true posterior density. Hence, $95\%$ credible intervals can be created using the variational posterior distribution for all the component means $\boldsymbol{\mu}_g$ and the component precision parameter ${1}/{\sigma_g^2}$ for each run (see\ Figure \ref{mean_tau}). A Bayesian credible interval provides an interval within which the unobserved parameter value falls with a certain probability. Similar to \cite{wang2005}, we also evaluated the frequentist coverage probability of the intervals, i.e., the number of times the true value of the parameter is contained within the credible interval. Across the nine parameters, the mean coverage probability was 0.927 (range 0.860--0.976), which is slightly lower than 0.95. \cite{blei2017} point out that variational inference tends to underestimate the variance of the posterior density.

For completeness, the EM algorithm together with the BIC to select the model (i.e., covariance structure and $G$)---referred to as the EM-BIC framework hereafter---was also applied to these data using the {\tt mclust} package for {\sf R}.  In all 50 cases, a $G=3$ component VII model is chosen and gives perfect classification results for all 50 datasets.

\subsection{Simulation Study 2}
We ran another simulation study with 50 different three component, three-dimensional Gaussian distributions with known mean and covariance structure $\matsig_g=\matsig=\lambda \matd \mata \matd'$.  Again, five different runs with different random starts are used and the maximum number of components is set to ten. In 41 out of 50 datasets, a three-component model is selected by the VB-DIC approach. Out of these 41 cases, an EEE model is selected 39 times and an EEV model is selected twice. These 41 cases give an average ARI of 1.000 (sd 0.001). When an EEV model is selected, the difference in DIC between the EEV and EEE models is 3.256 and 8.095, respectively, indicating that these two models were close in their fits. Four- and five-component models were selected for 8 and 1 of the datasets, respectively, with an average ARI of 0.923 (sd 0.097). The true and estimated mean parameters using VB-DIC for the EEE model are given in Table~\ref{table9}, and the true and estimated covariance parameters using VB-DIC for the EEE model are:
\begin{equation*}
\matsig =
\begin{bmatrix}
0.50& 0.35& 0.25 \\
0.35& 1.00 &0.45\\
0.25 &0.45& 1.20
\end{bmatrix}, \quad
\hat{\matsig} =
\begin{bmatrix}
0.494 ~(\text{sd}~0.049)& 0.346~(\text{sd}~0.044)& 0.235~(\text{sd}~0.046)\\
0.346 ~(\text{sd}~0.044)& 0.995~(\text{sd}~0.076)& 0.445~(\text{sd}~0.069)\\
0.235 ~(\text{sd}~0.046)&0.445~(\text{sd}~0.069)& 1.204~(\text{sd}~0.099)
\end{bmatrix}.
\end{equation*}
\begin{table}[ht]
\centering
\caption{Summary of the average and standard errors of the estimated parameters from 39 out of the 50 three-dimensional simulated datasets where an EEE model was selected along the true parameters used to generate the data.\label{table9}}
\begin{tabular*}{0.9\textwidth}{@{\extracolsep{\fill}}llccc}
\hline 
&&& \multicolumn{2}{c}{$\hat{\Mu}_g$} \\
\cline{4-5}
$g$&$n_g$&$\Mu_g$ & Mean&St.\ Dev. \\
\hline
1&150&$(-2,-2,-2)'$&$(-2.007, -2.025, -2.002)'$&$(0.077, 0.103, 0.119)'$\\
2&100&$(4,0,0)'$&$(4.002, 0.013, 0.017)'$&$(0.051, 0.068, 0.076)'$\\
3&75&$(-5,0,2)'$& $(-5.005, -0.009, 1.976)'$& $(0.087, 0.117, 0.108)'$\\
\hline
\end{tabular*}
\end{table}

The EM-BIC framework, via {\tt mclust}, was also used for these data. An EEE model was chosen for all 50 datasets with an average ARI of 1.0 (sd 0.001).

\subsection{Clustering of Benchmark Datasets}
To demonstrate the performance of the VB-DIC approach, we applied our algorithm on several benchmark datasets and compared its performance with the widely used EM-BIC framework via the {\tt mclust} package.\\[-6pt]

\begin{figure}[ht]
\begin{center}
\includegraphics[width=0.46\textwidth]{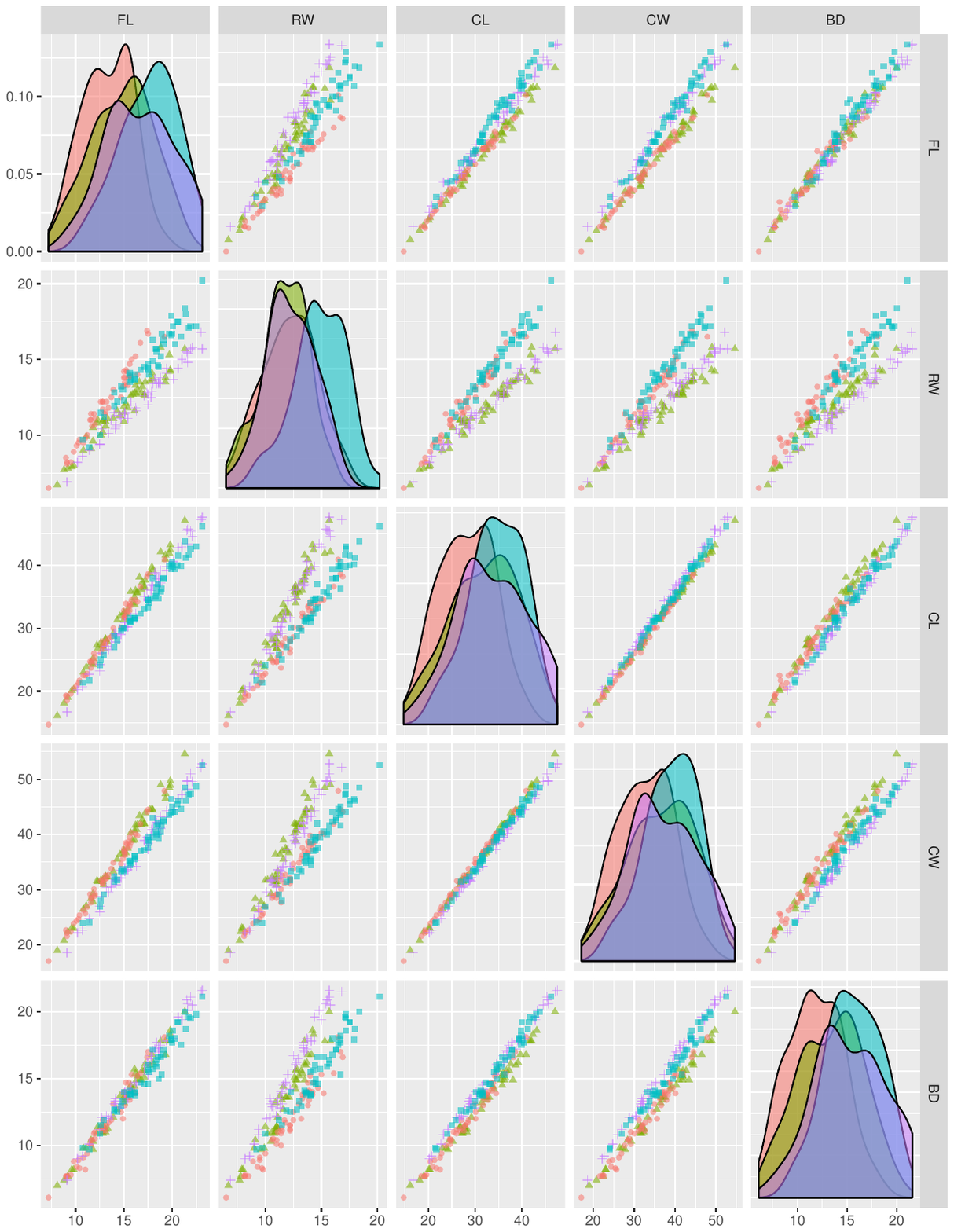} \quad
\includegraphics[width=0.46\textwidth]{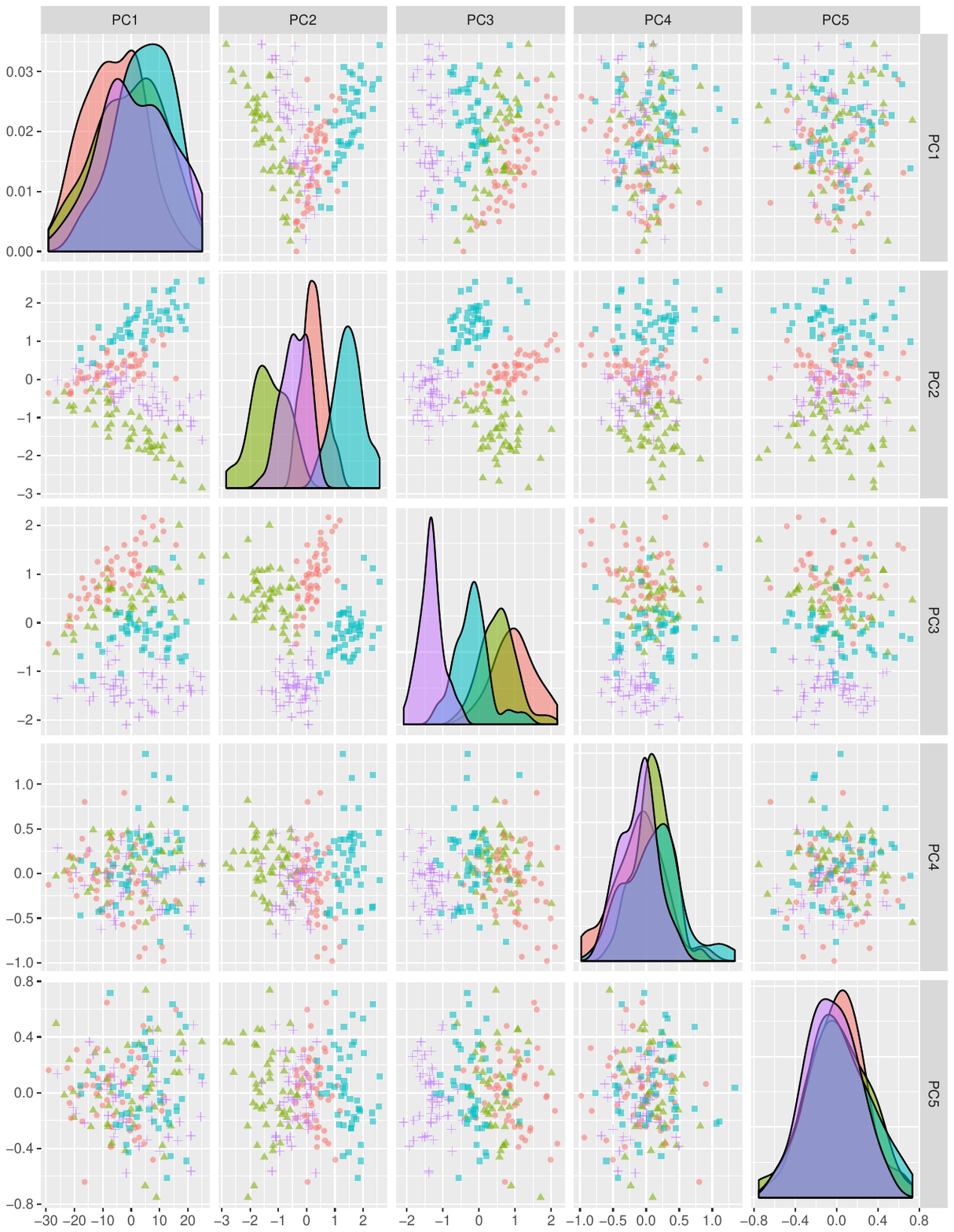} 
\caption{Scatter plot matrix showing the relationships among the variables in the Leptograpsus crabs dataset (left) and showing the relationships among the uncorrelated principal components (right), where the colors/symbols represents the different groups.}
\label{crabplot}
\end{center}
\end{figure}

\noindent \underline{Crabs data}: The \textit{Leptograpsus} crab data set, publicly available in the package {\tt MASS} \citep{venables2002} for {\sf R}, consists of biological measurements on 100 crabs from two different species (orange and blue) with 50 males and 50 females of each species. The biological measurements (in millimeters) include frontal lobe size, rear width, carapace length, carapace width, and body depth. Although this data set has been analyzed quite often in the literature, using several different clustering approaches, the correlation among the variables makes it difficult to cluster (Figure~\ref{crabplot}). Due to this known issue with the data set, we introduced an initial step of processing using principal component analysis to convert these correlated variables into principal components (Figure~\ref{crabplot}). Finally, the VB-DIC approach was run on these uncorrelated principal components with a maximum of $G=10$ components.\\[-6pt]

\noindent \underline{SRBCT data}: The \texttt{SRBCT} dataset, available in the {\sf R} package \texttt{plsgenomics} \citep{Boulesteix2018}, is a gene expression data from the microarray experiments of small round blue cell tumors (SRBCT) of childhood cancer. It contains measurements on 2,308 genes from 83 samples comprising of 29 cases of Ewing sarcoma (EWS), 11 cases of Burkitt lymphoma (BL), 18 cases of neuroblastoma (NB), and 25 cases of rhabdomyosarcoma (RMS). Note that our proposed variational Bayes algorithm is not designed for high-dimensional, low sample size (i.e., large $p$, small $N$) problems. \cite{dang2015} performed a differential expression analysis on the gene expression data using an ANOVA across the known groups and selected the top ten genes, ranked using the obtained p-values, to represent a potential set of measurements that contain
information on group identification. Hence, we preprocessed the \texttt{SRBCT} dataset in a similar manner to \cite{dang2015} and implemented the VB-DIC approach with a maximum of $G=10$ components.\\[-6pt]

\noindent \underline{Iris data}: The \texttt{Iris} data set available in the {\sf R} datasets package contains measurements in centimeters of the variables \texttt{sepal length, sepal width, petal length,} and \text{petal width} of 50 flowers from each of the three species of Iris: \textit{Iris setosa}, \textit{Iris versicolor}, and \textit{Iris virginica}.\\[-6pt]

\noindent \underline{Diabetes data}: The \texttt{diabetes} dataset available in the R package \texttt{mclust} contains measurements on three variables on 145 non-obese adult patients classified into three groups (Normal, Overt and Chemical): 
\begin{itemize}[noitemsep,leftmargin=*,topsep=0pt] 
\item \texttt{glucose}: Area under plasma glucose curve after a three hour oral glucose tolerance test.
\item \texttt{insulin}: Area under plasma insulin curve after a three hour oral glucose tolerance test.
\item \texttt{sspg}: Steady state plasma glucose.\\[-6pt]
\end{itemize}

\noindent \underline{Banknote data}: The \texttt{banknote} dataset, available in the {\sf R} package \texttt{mclust}, contains six measurements of 100 genuine and 100 counterfeit old Swiss 1000-franc bank notes. Measurements are available for the following variables:
\begin{itemize}[noitemsep,leftmargin=*,topsep=0pt] 
\item \texttt{Length}: Length of the bill in mm.
\item \texttt{Left}: Width of left edge in mm.
\item \texttt{Right}: Width of right edge in mm.
\item \texttt{Bottom}: Bottom margin width in mm.
\item \texttt{Top}: Top margin width in mm.
\item \texttt{Diagonal}: Length of diagonal in mm.
\end{itemize}

A summary of the performance of the VB-DIC approach and the EM-BIC approach is given in Table~\ref{tabrealdata}, where the approach that gives the best performance is in bold. For three out of five benchmark datasets, our VB-DIC approach outperforms the EM-BIC framework as implemented via \texttt{mclust}. For one of the five datasets, our VB-DIC approach gives the same ARI as the EM-BC framework and, in the other one of the five datasets, the EM-BIC framework yields a slightly larger ARI compared to our VB-DIC approach.\begin{table}[ht]
\centering
\caption{Summary of the performance of VB-DIC approach on the benchmark datasets along with the performance using the EM-BIC framework.\label{tabrealdata}}
\begin{tabular*}{0.95\textwidth}{@{\extracolsep{\fill}}lrcrcr}
\hline 
 &  &  \multicolumn{2}{c}{VB-DIC}& \multicolumn{2}{c}{EM-BIC}\\
 \cline{3-4}
  \cline{5-6}
 Dataset &Classes  &  $G$ & ARI& $G$ & ARI\\
\hline
\texttt{Banknote}&2&3&\textbf{0.842}&3&\textbf{0.842}\\
\texttt{Crabs}&4&5&\textbf{0.788}&6&0.600\\
\texttt{Diabetes}&3&4&0.645&3&\textbf{0.664}\\
\texttt{Iris}&3&3&\textbf{0.732}&2&0.568\\
\texttt{SRBCT}&4&4&\textbf{0.760}&4&0.736\\
\hline
\end{tabular*}
\end{table}

\section{Discussion}
A variational Bayes approach has been introduced for parameter estimation for model-based clustering via the well-known GPCM family. As stated before, an advantage of using a variational Bayes algorithm is that because the hyperparameters of the approximating posterior densities are updated at every iteration, we are indeed updating the approximating variational posterior density of a parameter as opposed to the point estimate of a parameter as in an EM framework. This essentially leads to a natural framework to extract interval estimates (i.e., credible intervals) for every run similar to a fully Bayesian approach but without the need to create a confidence interval via bootstrapping like in an EM framework. We also preserve the monotonicity property of the log-likelihood function like an EM algorithm which is lost in a fully Bayesian MCMC based-approach. Additionally, the variational Bayes approach allows for simultaneously obtaining parameter estimates and the number of components. However, a model selection criterion still needs to be utilized while selecting the covariance structure. Herein, we used the DIC for the selection of the covariance structure and so the resulting variational Bayes approach was called the VB-DIC approach. As can be seen from the simulation studies, the correct covariance structure is often selected using the DIC. That said, it may well be the case that another criterion is more suitable for selecting the model (i.e., the covariance structure). Notably, starting values play a different role for variational Bayes than for the EM algorithm---because the former gradually reduces $G$ as the algorithm iterates, the ``starting values'' for all but the initial $G$ are not the values used to actually start the algorithm. Accordingly, direct comparison of the VB-DIC and EM-BIC approaches is not entirely straightforward.

In the simulation studies, the parameters estimated using variational Bayes approximations were very close to the true parameters (when the correct model was chosen), and excellent classification was obtained using the model selected by DIC. In many of the simulated and real analyses, the performance of the VB-DIC approach was very similar or the same as the EM-BIC approach. This is not surprising. As noted by \cite{mcLachlan2008} and \cite{gelman2013}, the EM algorithm can be thought of as a special case of variational Bayes in which the parameters are partitioned into two parts, $\phi$ and $\gamma$, and where the approximating distribution of $\phi$ is required to be a point mass and the approximating distribution of $\gamma$ is unconstrained conditional on the last update of $\phi$. Across all the simulations, EM-BIC framework outperformed the VB-DIC approach; however, the VB-DIC approach outperformed the EM-BIC framework on three of the five benchmark real datasets considered.

In summary, we have explored a Bayesian alternative to the most widely used family of Gaussian mixture models, i.e., the GPCM family. The use of variational Bayes in conjunction with the DIC for a family of mixture models is a novel idea and lends itself nicely to further research. Moreover, the DIC provides an alternative model selection criterion to the almost ubiquitous BIC. There are several possible avenues for further research, one of which is extension to the semi-supervised \citep[e.g.,][]{mcnicholas10c} or, more generally, fractionally supervised paradigm \citep[see][]{vrbik15,gallaugher19}. Another avenue is extension to other families of Gaussian mixture models \citep[e.g., the PGMM family of][]{mcnicholas08,mcnicholas10d} and to non-Gaussian families of mixture models \citep[e.g.,][]{vrbik14,lin14}. Further consideration is needed vis-\`{a}-vis the approach used to selected the model (i.e., covariance structure) in the variational Bayes approach, e.g., one could conduct a detailed comparison of VB-DIC and, \textit{inter alia}, VB-BIC. Finally, an analogous variational Bayes approach could be taken to parameter estimation for mixtures of matrix variate distributions \citep[see, e.g.,][]{viroli11,gallaugher18a,gallaugher18b,gallaugher19b}.

{\small
\subsection*{Acknowledgements}

This work was supported by a Postgraduate Scholarship from the Natural Science and Engineering Research Council of Canada (Subedi); the Canada Research Chairs program (McNicholas); and an E.W.R.~Steacie Memorial Fellowship (McNicholas).


}

\newpage
\appendix
\section{Posterior distributions for the parameters of eigen-decomposed covariance matrix}\label{app:A}

{\small
\begin{longtable}{ p{.07\textwidth}  p{.275\textwidth} p{.65\textwidth}} 
\caption[]{Posterior distributions of the precision parameters as well as their corresponding parameters for 12 of the members of the GPCM family.\label{table3}}\\[-10pt]
\hline 
Model & Posterior Distributions & Parameters\\
\hline
\endfirsthead

\hline 
Model & Posterior Distributions & Parameters\\
\hline 
\endhead
 &&Continued on next page...\\ \hline
\endfoot

\endlastfoot
EII & Gamma $(a,b)$ & $a=a^{(0)}+d\sum_{g=1}^{G}\sum_{i=1}^{n}\hat{z}_{ig}=a^{(0)}+d n$\\
&&$b=b^{(0)}+\sum_{g=1}^G(\sum_{i=1}^{n}\hat{z}_{ig}\vecy_i'\vecy_i+\beta_g^{(0)}\mathbf{m}_g^{(0)T}\mathbf{m^{(0)}}- \beta_g\mathbf{m}_g'\mathbf{m})$\\
\hline
VII&Gamma $(a_g,b_g)$ &$a_g=a_g^{(0)}+d\sum_{i=1}^{n}\hat{z}_{ig}$\\
&&$b_g=b_g^{(0)}+\sum_{i=1}^{n}\hat{z}_{ig}\vecy_i'\vecy_i+\beta_g^{(0)}\mathbf{m}_g^{(0)T}\mathbf{m^{(0)}}- \beta_g\mathbf{m}_g'\mathbf{m}$\\
\hline
EEI&Gamma $(a_k,b_k)$ & $a_k=a_k^{(0)}+\sum_{g=1}^G \sum_{i=1}^n\hat{z}_{ig}=a_k^{(0)}+n$\\
&&$b_k=b_k^{(0)}+\sum_{g=1}^G\sum_{i=1}^{n}(\hat{z}_{ig}y_{ik}^2+\beta_g^{(0)}m_{gk}^{(0)2}- \beta_gm_{gk}^2)$\\
\hline
VEI &Gamma $(a_g,b_g)$ &$a_g=a_g^{(0)}+d\sum_{i=1}^{n}\hat{z}_{ig}$\\
&&$b_g=b_g^{(0)}+\sum_{i=1}^{n}\hat{z}_{ig}\vecy_i'\vecy_i+\beta_g^{(0)}\mathbf{m}_g^{(0)T}\mathbf{m^{(0)}}- \beta_g\mathbf{m}_g'\mathbf{m}$\\
&Gamma $(al_k,be_k)$& $al_k=al_k^{(0)}+\sum_{g=1}^G \sum_{i=1}^n\hat{z}_{ig}=al_k^{(0)}+n$\\
&&$be_k=be_k^{(0)}+\sum_{g=1}^G\sum_{i=1}^{n}(\hat{z}_{ig}y_{ik}^2+\beta_g^{(0)}m_{gk}^{(0)2}- \beta_gm_{gk}^2)$\\
\hline
EVI &Gamma $(a,b)$ &$a=a^{(0)}+d\sum_{g=1}^{G}\sum_{i=1}^{n}\hat{z}_{ig}=a^{(0)}+dn$ \\
&&$b=b^{(0)}+\sum_{g=1}^G\sum_{i=1}^{n}\hat{z}_{ig}\vecy_i'\vecy_i+\beta_g^{(0)}\mathbf{m}_g^{(0)T}\mathbf{m^{(0)}}- \beta_g\mathbf{m}_g'\mathbf{m}$\\
&  Gamma $(al_{gk},be_{gk})$& $al_{gk}=al_{gk}^{(0)}+ \sum_{i=1}^n\hat{z}_{ig}$\\
&&$be_{gk}=be_{gk}^{(0)}+\sum_{i=1}^{n}\hat{z}_{ig}y_{ik}^2+\beta_g^{(0)}m_{gk}^{(0)2}- \beta_gm_{gk}^2$\\
\hline
VVI &Gamma $(a_g,b_g)$ &$a_g=a_g^{(0)}+d\sum_{i=1}^{n}\hat{z}_{ig}$\\
&&$b_g=b_g^{(0)}+\sum_{i=1}^{n}\hat{z}_{ig}\vecy_i'\vecy_i+\beta_g^{(0)}\mathbf{m}_g^{(0)T}\mathbf{m^{(0)}}- \beta_g\mathbf{m}_g'\mathbf{m}$\\
&Gamma $(al_{gk},be_{gk})$ &$al_{gk}=al_{gk}^{(0)}+\sum_{i=1}^n\hat{z}_{ig}$\\
&&$be_{gk}=be_{gk}^{(0)}+\sum_{i=1}^{n}\hat{z}_{ig}y_{ik}^2+\beta_g^{(0)}m_{gk}^{(0)2}- \beta_gm_{gk}^{2}$\\
\hline
EEE & Wishart $(v,\Sigma^{-1})$ & $v = v^{(0)}+\sum_{g=1}^G\sum_{i=1}^n\hat{z}_{ig}= v^{(0)}+n$\\
&&$\Sigma^{-1}=\Sigma^{(0)-1}+\sum_{g=1}^G(\sum_{i=1}^{n}\hat{z}_{ig}\vecy_i'\vecy_i+\beta_g^{(0)}\mathbf{m}_g^{(0)T}\mathbf{m^{(0)}}- \beta_g\mathbf{m}_g'\mathbf{m})$\\
\hline
VEE & Gamma $(a_g,b_g)$ &$a_g=a_g^{(0)}+d\sum_{i=1}^{n}\hat{z}_{ig}$\\
&&$b_g=b_g^{(0)}+\sum_{i=1}^{n}\hat{z}_{ig}\vecy_i'\vecy_i+\beta_g^{(0)}\mathbf{m}_g^{(0)T}\mathbf{m^{(0)}}- \beta_g\mathbf{m}_g'\mathbf{m}$\\
& Wishart $(v,\Sigma)$ &$v = v^{(0)}+\sum_{g=1}^G\sum_{i=1}^n\hat{z}_{ig}= v^{(0)}+n$\\
&&$\Sigma=\Sigma^{(0)}+\sum_{g=1}^G(\sum_{i=1}^{n}\hat{z}_{ig}\vecy_i'\vecy_i+\beta_g^{(0)}\mathbf{m}_g^{(0)T}\mathbf{m^{(0)}}- \beta_g\mathbf{m}_g'\mathbf{m})$\\
\hline
EEV& Gamma $(a_k,b_k)$ & $a_k=a_k^{(0)}+d\sum_{g=1}^{G}\sum_{i=1}^{n}\hat{z}_{ig}=a_k^{(0)}+dn$ \\
&&$b_k=b_k^{(0)}+\sum_{g=1}^G(\sum_{i=1}^{n}\hat{z}_{ig}y_{ik}^2+\beta_g^{(0)}m_{gk}^2- \beta_gm_{gk}^{2})$\\
&  Bingham matrix $(P_g,Q)$& See mathematical details for the EEV model below.\\
\hline
VEV &Gamma $(a_g,b_g)$ & $a_g=a_g^{(0)}+d\sum_{i=1}^{n}\hat{z}_{ig}$\\
&&$b_g=b_g^{(0)}+\sum_{i=1}^{n}\hat{z}_{ig}\vecy_i'\vecy_i+\beta_g^{(0)}\mathbf{m}_g^{(0)T}\mathbf{m^{(0)}}- \beta_g\mathbf{m}_g'\mathbf{m}$\\ 
&Gamma $(al_k,be_k)$& $al_k=al_k^{(0)}+\sum_{g=1}^G \sum_{i=1}^n\hat{z}_{ig}=al_k^{(0)}+n$\\
&&$be_k=be_k^{(0)}+\sum_{g=1}^G(\sum_{i=1}^{n}\hat{z}_{ig}y_{ik}^2+\beta_g^{(0)}m_{gk}^{(0)2}- \beta_gm_{gk}^{2})$\\
& Bingham matrix ($P_g,Q_g$) &See posterior for $\matd_g$ in the VEV Model below.\\
\hline
EVV& Gamma $(a,b)$& $a=a^{(0)}+d\sum_{g=1}^{G}\sum_{i=1}^{n}\hat{z}_{ig}=a^{(0)}+dn$ \\
&&$b=b^{(0)}+\sum_{g=1}^G(\sum_{i=1}^{n}\hat{z}_{ig}\vecy_i'\vecy_i+\beta_g^{(0)}\mathbf{m}_g^{(0)T}\mathbf{m^{(0)}}- \beta_g\mathbf{m}_g'\mathbf{m})$\\
 &Wishart $(v_g,\Sigma_g^{-1})$ & $v_g = v_g^{(0)}+\sum_{i=1}^n\hat{z}_{ig}$\\
&&$\Sigma_g^{-1}=\Sigma_g^{(0)-1}+\sum_{i=1}^{n}\hat{z}_{ig}\vecy_i\vecy_i'+\beta_g^{(0)}\mathbf{m_g^{(0)}}\mathbf{m}_g^{(0)T}- \beta_g\mathbf{m}_g\mathbf{m}_g'$\\
\hline
VVV& Wishart $(v_g,\Sigma_g^{-1})$ & $v_g = v_g^{(0)}+\sum_{i=1}^n\hat{z}_{ig}$\\
&&$\Sigma_g^{-1}=\Sigma_g^{(0)-1}+\sum_{i=1}^{n}\hat{z}_{ig}\vecy_i\vecy_i'+\beta_g^{(0)}\mathbf{m_g^{(0)}}\mathbf{m}_g^{(0)T}- \beta_g\mathbf{m}_g\mathbf{m}_g'$\\
\hline

\end{longtable}
}
\newpage

\section{Posterior expected value of the precision parameters of the eigen-decomposed covariance matrix}

{\small
\begin{longtable}{ p{.10\textwidth}  p{.12\textwidth} p{.68\textwidth}} 
\caption{Posterior expected value of the precision parameters of the eigen-decomposed covariance matrix for 12 of the members of the GPCM family.}\\[-10pt]
\hline 
Model & Parameters & Expected Values\\
\hline
\endfirsthead
\hline 
Model & Parameters & Expected Values\\
\hline 
\endhead
 &&Continued on next page...\\ \hline
\endfoot
\endlastfoot
EII & $\lambda \ident_d$ & $\ex[(\lambda)^{-1}]={a}/{b}$ \\
& &$\ex[\log|(\lambda)^{-1}|]=\Psi({a}/{2})-\log({b}/{2})$\\
\hline
VII &$\lambda_g \ident_d$  & $\ex[(\lambda_g)^{-1}]={a_g}/{b_g}$ \\
&& $\ex[\log|(\lambda_g)^{-1}|]=\Psi({a_g}/{2})-\log({b_g}/{2})$\\
\hline
EEI&$\lambda \mata$ & $\ex[(\lambda\mata)^{-1}_{k,k}]={a_k}/{b_k}$ \\
& &$\ex[\log|(\lambda \mata)^{-1}_{k,k}|]=\Psi({a_k}/{2})-\log({b_k}/{2})$\\
\hline
VEI &$\lambda_g \mata$&$\ex[\lambda^{-1}_g]={a_g}/{b_g}$ \\
&& $\ex[\log|\lambda^{-1}_g|]=\Psi({a_g}/{2})-\log({b_g}/{2})$\\
&& $\ex[(c\mata^{-1})_{k,k}]={(al_k)}/{(be_k)}$ \\
& &$\ex[\log|(c\mata^{-1})_{k,k}|]=\Psi({al_k}/{2})-\log({be_k}/{2})$\\
\hline
EVI &$\lambda \mata_g$&$\ex[\lambda^{-1}]={a}/{b}$ \\
&& $\ex[\log|\lambda^{-1}|]=\Psi({a}/{2})-\log({b}/{2})$\\
&& $\ex[(c_g\mata^{-1}_g)_{k,k}]={a_{gk}}/{b_{gk}}$ \\
& &$\ex[\log|(c_g\mata^{-1}_g)_{k,k}|]=\Psi({a_{gk}}/{2})-\log({b_{gk}}/{2})$\\
\hline
VVI &$\lambda_g \mata_g$ &$\ex[\lambda^{-1}_g]={a_g}/{b_g}$ \\
&& $\ex[\log|\lambda^{-1}_g|]=\Psi({a_g}/{2})-\log({b_g}/{2})$\\
&& $\ex[(c_g\mata^{-1}_g)_{k,k}]={a_{gk}}/{b_{gk}}$\\
& &$\ex[\log|(c_g\mata^{-1}_g)_{k,k}|]=\Psi({a_{gk}}/{2})-\log({b_{gk}}/{2})$\\
\hline
EEE & $\lambda \matd \mata \matd'$ & $\ex[(\lambda \matd \mata \matd')^{-1}]=v\matsig^{-1}$\\
&&$\ex[\log|(\lambda \matd \mata \matd')^{-1}|]=\sum_{k=1}^d\Psi({(v+1-k)}/{2})+d\log (2)-\log |\matsig|$\\
\hline
VEE & $\lambda_g \matd \mata \matd'$ & $\ex[\lambda^{-1}_g]={a_g}/{b_g}$ \\
&& $\ex[\log|\lambda^{-1}_g|]=\Psi({a_g}/{2})-\log({b_g}/{2})$\\
&&$\ex[( \matd \mata \matd')^{-1}]=v\matsig^{-1}$\\
&&$\ex[\log|( \matd \mata \matd')^{-1}|]=\sum_{k=1}^d\Psi({(v+1-k)}/{2})+d\log (2)-\log |\matsig|$\\
\hline
EEV& $\lambda \matd_g\mata \matd_g'$  & $\ex[(\lambda \mata)_{k,k}]={a_k}/{b_k}$ \\
& &$\ex[\log|(\lambda \mata)_{k,k}|]=\Psi({a_k}/{2})-\log({b_k}/{2})$\\
& &$\ex[(\lambda \matd_g\mata \matd_g')^{-1}|(\lambda \mata)^{-1}] \text{ via Monte Carlo integration}$\\
\hline
VEV &$\lambda_g \matd_g\mata \matd_g'$ & $\ex[\lambda^{-1}_g]={a_g}/{b_g}$ \\
&& $\ex[\log|\lambda^{-1}_g|]=\Psi({a_g}/{2})-\log({b_g}/{2})$\\
&& $\ex[(c\mata^{-1})_{k,k}]={(al_k)}/{(be_k)}$ \\
& &$\ex[\log|(\mata^{-1})_{k,k}|]=\Psi({al_k}/{2})-\log({be_k}/{2})$\\
& &$\ex[(\lambda_g^{-1} \matd_g\mata\matd_g')^{-1}|(\lambda_g \mata)^{-1}] \text{ via Monte Carlo integration}$\\
\hline
EVV& $\lambda \matd_g\mata_g \matd_g'$& $\ex[\lambda]={a}/{b}$ \\
& &$\ex[\log\lambda]=\Psi({a}/{2})-\log({b}/{2})$\\
&&$\ex[(\matd_g\mata_g \matd_g')^{-1}]=v\matsig_g^{-1}$\\
&&$\ex[\log|(\matd_g\mata_g \matd_g')^{-1}|]=\sum_{k=1}^d\Psi({(v_g+1-k)}/{2})+d\log (2)-\log |\matsig_g|$\\
\hline
VVV& $\lambda_g \matd_g\mata_g \matd_g'$ & $\ex[(\matd_g\mata_g \matd_g')^{-1}]=v\matsig_g^{-1}$\\
&&$\ex[\log|(\matd_g\mata_g \matd_g')^{-1}|]=\sum_{k=1}^d\Psi({(v_g+1-k)}/{2})+d\log (2)-\log |\matsig_g|$\\
\hline

\label{tab:myfirstlongtable}
\end{longtable}
}

\section{Mathematical Details for the EEV and VEV Models}\label{app:math}
\subsection{EEV Model}
The mixing proportions were assigned a Dirichlet prior distribution, such that 
$$q_{\rho}(\boldsymbol{\rho})=\text{Dir}(\boldsymbol{\rho};\alpha_1^{(0)},\ldots,\alpha_G^{(0)}).$$
For the mean, a Gaussian distribution conditional on the covariance matrix was used, such that
\begin{equation*}
q_{\Mu}(\Mu \mid\lambda,\mata,\matd_1,\ldots,\matd_G)=\prod_{g=1}^{G}\phi_{d}(\Mu_g;\mathbf{m}_g^{(0)},(\beta_g^{(0)-1}\lambda\matd_g\mata\matd_g')).
\end{equation*}
For the parameters of the covariance matrix, the following priors were used:
the $k$th diagonal elements of $(\lambda\mata)^{-1}$ were assigned a Gamma $(a^{(0)}_k,b^{(0)}_k)$ distribution and $\matd_g$ was assigned a matrix von~Mises-Fisher $(\mathbf{C}^{(0)}_g)$ distribution. By setting $\boldsymbol{\tau} = (\lambda\mata)^{-1}$, its prior can be written $$p_{\tau}(\boldsymbol{\tau})\propto \prod_{k=1}^K\tau_{k}^{\frac{a^{(0)}_k}{2}-1} \exp\left\{-\frac{b^{(0)}_k}{2}\tau_k\right\},$$
where $\tau_k$ is the $k$th diagonal element of $\boldsymbol{\tau}=(\lambda\mata)^{-1}$.

The matrix $\matd$ has a density as defined by \cite{gupta2000}:
$$p(\matd) =  b(\mathbf{Q}^{(0)},\mathbf{P}_g^{(0)})\exp(\tr\{\mathbf{Q}^{(0)}\matd\mathbf{P}_g^{(0)}\matd'\}) [d\matd],$$
for $\matd \in O(d,d)$, where $O(d,d)$ is the Stiefel manifold of $d\times d$ matrices, $[d\matd]$ is the unit invariant measure on $O(d,d)$, and $\mata^{(0)}$ and $\mathbf{B}_g^{(0)}$ are symmetric and diagonal matrices, respectively.

The joint distribution of $\Mu_1,\ldots,\Mu_G$, $\boldsymbol{\tau}$, and $\matd$ is
\begin{equation*}\begin{split}
p(\Mu_1,\ldots,\Mu_G,\boldsymbol{\tau}, \matd) \propto\prod_{g=1}^G&|\beta_g^{(0)}\boldsymbol{\tau}|^{\frac{1}{2}} \exp \left\{\frac{-(\Mu_g-\mathbf{m}_g^{(0)})\beta_g^{(0)}\matd_g'\boldsymbol{\tau} \matd_g(\Mu_g-\mathbf{m}_g^{(0)})'}{2} \right\}\\ 
&\qquad\times\exp \left\{\tr(\mathbf{Q}^{(0)}\matd\mathbf{P}_g^{(0)}\matd')\right\}\prod_{k=1}^K\tau_{k}^{\frac{a^{(0)}_k}{2}-1}
\exp\left\{-\frac{b^{(0)}_k}{2}\tau_k\right\}.
\end{split}
\end{equation*}
The likelihood of the data can be written
\begin{equation*}
\mathcal{L}(\Mu_1,\ldots,\Mu_G,\boldsymbol{\tau}, \matd \mid \vecy_1, \ldots, \vecy_n) \propto\prod_{i=1}^{n}\prod_{g=1}^{G}|\boldsymbol{\tau}|^{{\hat{z}_{ig}}/{2}} \exp \left\{-\frac{\hat{z}_{ig}}{2}(\vecy_i-\Mu_g)\matd_g'\boldsymbol{\tau} \matd_g(\vecy_i-\Mu_g)'\right\}.\end{equation*}
Therefore, the joint posterior distribution of $\Mu$, $\boldsymbol{\tau}$, and $\matd$ can be written
$$p(\Mu_1,\ldots,\Mu_G,\boldsymbol{\tau}, \matd \mid \vecy_1, \ldots, \vecy_n) \propto p(\Mu_1,\ldots,\Mu_G,\boldsymbol{\tau}, \matd) \mathcal{L}(\Mu_1,\ldots,\Mu_G,\boldsymbol{\tau}, \matd \mid \vecy_1, \ldots, \vecy_n).$$
Thus, the posterior distribution of mean becomes $$q_{\Mu}(\Mu_1,\ldots,\Mu_G \mid \boldsymbol{\tau},\matd_1,\ldots,\matd_G)=\prod_{g=1}^{G}\phi_d(\Mu_g;\mathbf{m}_g,(\beta_g\matd_g'\boldsymbol{\tau}\matd_g)^{-1}),$$
where $\beta_g = \beta_g^{(0)}+\sum_{i=1}^n\hat{z}_{ig}$ and
$$\mathbf{m}_g =\frac{1}{\beta_g}\left(\beta_g^{(0)} \mathbf{m}_g^{(0)}+\sum_{i=1}^n\hat{z}_{ig}\mathbf{y}_i\right).$$

The posterior distribution for the $k$th diagonal element of $\boldsymbol{\tau}=(\lambda\mata)^{-1}$ is 
$$q_{\tau}(\tau_k) = \mbox{Gamma} (a_k,b_k)$$ where $a_k=a_k^{(0)}+d\sum_{g=1}^{G}\sum_{i=1}^{n}\hat{z}_{ig}=a_k^{(0)}+dn$ and $$b_k=b_k^{(0)}+\sum_{g=1}^G\left(\sum_{i=1}^{n}\hat{z}_{ig}y_{ik}^2+\beta_g^{(0)}m_{gk}^2- \beta_gm_{gk}^{2}\right).$$
We have
\begin{equation*}\begin{split} 
q(\matd_g|\vecy;\Mu_g,\boldsymbol{\tau})&\propto \exp\left\{ \tr \left (-\frac{1}{2}{(\Mu_g-\mathbf{m}_g^{(0)})\beta_g^{(0)}\matd_g'\boldsymbol{\tau} \matd_g(\Mu_g-\mathbf{m}_g^{(0)})'}\right ) \right \} \\
 &\times\exp\left\{ \tr \left (-\frac{1}{2}{\sum_{i=1}^{n}z_{ig}(\vecy-\Mu_g)\matd_g'\boldsymbol{\tau} \matd_g(\vecy-\Mu_g)'}+\mathbf{Q}_g^{(0)}\matd_g\mathbf{P}_g^{(0)}\matd_g'\right )\right \},
\end{split}
\end{equation*}
which has the functional form of a Bingham matrix distribution, i.e., the form
$$\exp \left\{{\tr(\mathbf{Q}_g \matd_g\mathbf{P}_g \matd_g'})\right\},$$
where $\mathbf{Q}_{g} = \mathbf{Q}_{g}^{(0)}+\boldsymbol{\tau}$ and
\begin{equation*}
\mathbf{P}_g = \mathbf{P}_g^{(0)}-\frac{1}{2}\left[\sum_{i=1}^nz_{ig}(\vecy-\Mu_g)(\vecy-\Mu_g)'+(\Mu_g-\mathbf{m}_g^{(0)})\beta_g^{(0)}(\Mu_g-\mathbf{m}_g^{(0)})'\right].
\end{equation*}

\subsection{VEV Model}
Similarly, the posterior distribution of $\matd_g$ for the VEV model has the form
\begin{equation*}\begin{split} 
q(\matd_g|\vecy;\Mu_g,\boldsymbol{\tau}_g)&\propto \exp\left\{ \tr \left (-\frac{1}{2}{(\Mu_g-\mathbf{m}_g^{(0)})\beta_g^{(0)}\matd_g'\boldsymbol{\tau}_g \matd_g(\Mu_g-\mathbf{m}_g^{(0)})'}\right ) \right \}  \\
 &\times \exp\left\{ \tr \left (-\frac{1}{2}\sum_{i=1}^n\hat{z}_{ig}(\vecy-\Mu_g)\matd_g'\boldsymbol{\tau}_g \matd_g(\vecy-\Mu_g)'+\mathbf{Q}_g^{(0)}\matd_g\mathbf{P}_g^{(0)}\matd_g'\right )\right \},
\end{split}\end{equation*}
which has the functional form of a Bingham matrix distribution, i.e., the form
$$\exp\left\{{\tr(\mathbf{Q}_g \matd_g\mathbf{P}_g \matd_g'})\right\},$$
where $\mathbf{Q}_g=\mathbf{Q}_{g}^{(0)}+\boldsymbol{\tau}_g$ and 
$$\mathbf{P}_g = -\frac{1}{2}\left[\sum_{i=1}^n\hat{z}_{ig}(\vecy-\Mu_g)(\vecy-\Mu_g)'+(\Mu_g-\mathbf{m}_g^{(0)})\beta_g^{(0)}(\Mu_g-\mathbf{m}_g^{(0)})'\right].$$

\end{document}